\documentclass[aps,prl,twocolumn]{revtex4}
\usepackage{amsfonts}
\usepackage{amsmath}
\usepackage{amssymb}
\usepackage{graphicx}
\usepackage{color}

\def\beq{\begin{equation}}
\def\eeq{\end{equation}}
\def\beqa{\begin{eqnarray}}
\def\eeqa{\end{eqnarray}}

\def\e{\epsilon}

\def\e{\varepsilon}

\def\etal{{\sl et al.}}

\def\nonum{\nonumber \\}

\def\nonum{ \nonumber \\}

\renewcommand{\sec}[1]{\vskip 0.truecm \noindent {\sl #1}. -- }

\newcommand{\bq}{\begin{equation}}
\newcommand{\eq}{\end{equation}}

\begin{document}

\preprint{}
\title{Tunneling into clean Heavy Fermion Compounds: Origin of the Fano
Lineshape }
\author{P. W\"{o}lfle$^{1}$, Y. Dubi$^{2}$ and A. V. Balatsky$^{2,3}$}

\affiliation{$^1$ Institute for Theory of Condensed Matter and Center for Functional Nanostructures, Karlsruhe Institute of Technology, D-76128 Karlsruhe, Germany}
\affiliation{$^2$ Theoretical Division, Los Alamos National Laboratory, Los Alamos, NM 87545, USA}
\affiliation{$^3$ Center for Integrated Nanotechnologies, Los Alamos National Laboratory, Los Alamos, NM 87545}

\date{\today }

\begin{abstract}
Recently observed tunneling spectra on clean heavy fermion compounds show a
lattice periodic Fano lineshape similar to what is observed in the case of
tunneling to a Kondo ion adsorbed at the surface. We show that the
translation symmetry of a clean surface in the case of \emph{weakly correlated}
metals leads to a tunneling spectrum given by the superposition of the local
weighted density of states of all energy bands involved, which does not have a Fano
lineshape. In particular the spectrum will show any hybridization gap
present in the band structure. By contrast, in a \emph{strongly correlated} heavy fermion metal the
heavy quasiparticle states will be broadened by interaction effects. The
broadening grows as one moves away from the Fermi surface, up to a value of
the order of $T_K$ , the Kondo scale. We show that the hybridization gap is completely filled in this way, and an ideal Fano lineshape of width $T_K$ results, similar to the impurity case. We also discuss the possible influence of the tunneling tip on the surface, in (i) leading to additional broadening of the Fano line, and (ii) enhancing the hybridization locally, hence adding to the impurity type behavior. The latter effects depend on the tip-surface distance.
\end{abstract}

\maketitle


\sec{Introduction} Recent progress in scanning tunneling spectroscopy (STS) techniques has made
it possible for the first time to measure the differential conductance,
dI/dV, in heavy-fermion compounds \cite{Schmidt,Aynajian}. These materials are governed by the
strong Coulomb interaction on the f-electron orbitals, hybridizing with a
conduction band. Provided the f-energy is sufficiently below the Fermi
energy and the hybridization strength is weak enough, magnetic moments will
appear at the f-sites. The Kondo effect will screen the local moments at a
temperature below the Kondo temperature $T_{K}$ \cite{Kondo,Hewson}. The ensuing Fermi liquid
state is characterized by large effective masse ratios $m^{\ast }/m$ . Out
of this heavy Fermi liquid a variety of interesting phenomena may emerge, such as magnetic
order, quantum phase transitions and the associated non-Fermi liquid
behavior, as well as unconventional superconductivity \cite{exp}. Their microscopic
origin likely lies in the competition between the exchange interaction of
the magnetic moments and their screening by conduction electrons \cite{Doniach}, though no
theoretical consensus has emerged as yet \cite{theory}. STS experiments, by providing
insight into the local electronic structure \cite{Schmidt,Aynajian} of heavy-fermion materials,
might hold the key to understanding their complex properties, by providing
complementary information. The theoretical challenge in the interpretation
of the differential conductance in Kondo lattice systems \cite{Maltseva,Morr} lies in the
proper treatment of the strong correlation effects governing these systems.
Whereas tunneling into single Kondo impurities \cite{Hershfield,Ujsaghy,Plihal,Merino,Madhavan,Nagaoka,Yang,Fogelstrom} is relatively well
understood, there is a serious problem of interpretation of tunneling into
states on a clean crystal surface, which has the lattice periodicity. In the
single impurity case the quantum interference between electrons tunneling
from the STS tip into the conduction band and into the magnetic f-electron
states is essential. As a consequence, the lineshape of the $dI/dV$ spectrum
shows an asymmetric form, as first discussed by Fano \cite{Fano}. It is rather
surprising that a similar spectrum is found, periodically continued along
the surface, in the clean case. Within the framework of electron band theory
one would rather expect the spectrum to show the superposition of local
density of states contributions from the different electronic bands
involved. Recent attempts to calculate the tunneling spectrum by treating
the problem of strong correlations within pseudoparticle mean field theory
bear out this expectation \cite{Maltseva,Morr}. The resulting spectra are
characterized by two peaks belonging to the two heavy bands, separated by a
hybridization gap. As we will show below, a mapping of the problem onto a
non-interacting system, as done in 1/N theories is not sufficient to capture
the dominant many-body effects at energy scales of the order of $T_{K}.$

In this paper, we address this issue within Fermi liquid theory extended to
somewhat higher energy scales. It is convenient to start with the lattice
Anderson model in the low energy domain where heavy quasiparticles (QP) are well
defined. Our main point is that the QP width $\Gamma $ at energy $E$, which
varies as $\Gamma $ $\propto E^{2}/T_{K}$ in the Fermi liquid regime (we use units with $\hbar=k_{B}=1$), will
grow up to $\Gamma $ $\propto T_{K}$ at the scale $E=T_{K}$, which is larger
than the hybridization gap, given by $\Delta \propto (V/\epsilon
_{F}^{0})T_{K}<<T_{K}$, where $V$ is the hybridization amplitude and $%
\epsilon _{F}^{0}$ is the bare Fermi energy of the unhybridized conduction
band. Consequently, the local f-electron density of states (DOS) takes the form of a Lorentzian of width $T_{K}$, just like in the impurity case, and the Fano lineshape
arises as usual. We will first consider the case of a weakly correlated
metal, to make our point that in this case no Fano lineshape is expected.

While the translation invariance is preserved in an ideal noninvasive tunneling experiment, in reality the translation invariance may be broken by the local influence the tip may exert on the surface. Such an effect has been recently invoked to explain the somewhat distorted energy spectrum observed in tunneling spectra of a topological insulator \cite{Cheng}. We discuss two ways by which the presence of the tip may change the tunneling spectra. Both effects will depend on the tip-surface distance.

\begin{figure}
\includegraphics[width=8.5truecm]{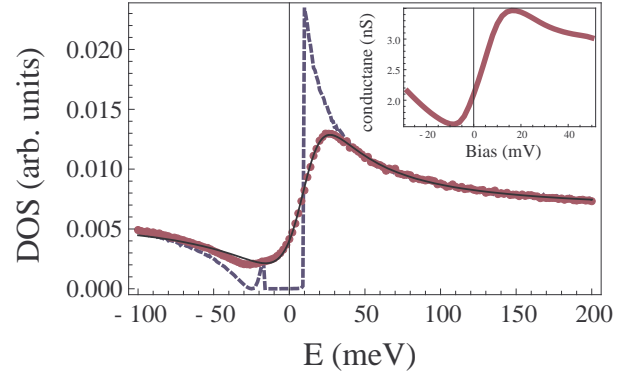}
\caption{Density of states for a two-band model of Eq.~\ref{HybridHamiltonian} without the QP width (dashed line) and with the QP width of Eq.~\ref{QPwidth} (solid line) with $T_K=130 $K (see text for other parameters). The inset shows the DOS with QP width (circles) and a fit to the Fano lineshape (solid line), yielding a Fano linewidth of $\sim 20$meV, in good agreement with experiment \cite{Schmidt,Aynajian}. For a qualitative comparison, in the inset we plot the experimental Fano lineshape, taken with permission from Ref.~\cite{Schmidt}.}
\label{figFano}
\end{figure}

\sec{Tunneling into weakly correlated metals} We consider tunneling into a metal characterized by two hybridized bands, as
described by the Hamiltonian
\begin{eqnarray}
H&=& H_0+ H_{hyb}+H_{t}\nonum
H_0&= &\sum_{\mathbf{k},\sigma }\epsilon _{\mathbf{k}\sigma }c_{%
\mathbf{k}\sigma }^{+}c_{\mathbf{k}\sigma }+\epsilon
_{f \sigma } \sum_{i,\sigma }n_{fi\sigma } \nonum
H_{hyb} &=&V\sum_{i,\mathbf{k},\sigma }(e^{i\mathbf{{k\cdot R}_{i}}%
}f_{i\sigma }^{+}c_{\mathbf{k}\sigma ^{\prime }}+h.c.)\nonum
H_{t} &=&t_{c}\sum_{\mathbf{k},\sigma }(p_{\sigma }^{+}c_{\mathbf{k}\sigma
}+h.c.)+t_{f}\sum_{\sigma }(p_{\sigma }^{+}f_{0\sigma }+h.c.) \nonum \label{HybridHamiltonian}
\end{eqnarray}%
where $c_{\mathbf{k}\sigma }^{+},f_{i\sigma }^{+},p_{\sigma }^{+}$ create an
electron of spin $\sigma $ in a Bloch state of momentum $\mathbf{k,}$ a
localized f-orbital at site $i$ or the level at the tip of the tunneling
electrode, respectively. The operator $n_{fi\sigma }=f_{i\sigma
}^{+}f_{i\sigma }$ counts the number of electrons with spin $\sigma $\ on
the local f-level, and $\epsilon_{f}$ is the energy of the f-level (which is position and momentum independent). In Eq.~(\ref{HybridHamiltonian}) we only take into account tunneling into the
orbitals directly under the tip at $\mathbf{{R}}_{0}=0$, which is the only source for breaking of the translational invariance, otherwise preserved in the Hybridized system. Assuming the
tunneling electrode and the metal to be in thermal equilibrium (in the limit
of vanishing tunneling current), their chemical potentials differing by $eV,$%
where $V$ is the applied bias voltage, the tunneling current to lowest order
in the tunneling amplitudes $t_{c},t_{f}$ (taken to be real valued) is given
by

\beqa
I(V)&=&\frac{2e}{\hbar }\int d\omega N_{t}(\omega -eV)[f(\omega -eV)-f(\omega )]\times \nonum ~&~& \times\Im
\{t_{c}^{2}G_{cc}(\omega )+t_{f}^{2}G_{ff}(\omega )+2t_{c}t_{f}G_{cf}(\omega
)\}
\eeqa %
where $G_{ab}(\omega )$ are the advanced local single particle Green's
functions of $H_{wc}=H_{0}+H_{hyb}$ and $N_t(\omega)$ is the STM tip DOS. The Hamiltonian $H_{wc}$ may easily be
diagonalized to yield two hybridized bands with Bloch energies

\begin{equation}
\epsilon _{\mathbf{k}}^{+,-}=\frac{1}{2}\left(\epsilon _{f}+\epsilon _{\mathbf{k}%
}\pm \sqrt{(\epsilon _{f}-\epsilon _{\mathbf{k}})^{2}+4V^{2}}\right)
\end{equation}%
and density of states

\begin{equation}
N_{+,-}(\omega )=\int d^{3}k\delta (\omega -\epsilon _{\mathbf{k}}^{+,-})
\end{equation}%
The density of states exhibits a hybridization gap at $\omega =\epsilon _{f}$
of width $\Delta =2V$ . The Green's functions are found as

\beqa
G_{ab}(\omega )&=&\sum\limits_{\nu =+,-}\frac{a_{ab,\mathbf{k}}^{\nu }}{%
\omega -\epsilon _{\mathbf{k}}^{\nu }},~\text{ \ \ }a_{ff,\mathbf{k}}^{\nu
}=\nu \frac{\epsilon _{\mathbf{k}}^{\nu }-\epsilon _{\mathbf{k}}}{2\Delta
\epsilon _{\mathbf{k}}}
\nonum \text{ \ }a_{cc,\mathbf{k}}^{\nu }&=&\nu \frac{%
\epsilon _{\mathbf{k}}^{\nu }-\epsilon _{f}}{2\Delta \epsilon _{\mathbf{k}}},%
\text{ \ }a_{cf,\mathbf{k}}^{\nu }=\nu \frac{V}{2\Delta \epsilon _{\mathbf{k}%
}}
\eeqa
where $\Delta \epsilon _{\mathbf{k}}=\epsilon _{\mathbf{k}}^{+}-\epsilon _{%
\mathbf{k}}^{-}$. While the coherence factors $a_{ab,\mathbf{k}}^{\nu }$
shift the weight in the partial spectral functions $\Im\{G_{ab}(\omega
)\}$ somewhat, compared to the band density of states $N_{+,-}(\omega ),$
the hybridization gap will remain. In Fig.~\ref{figFano}  we show the differential
conductance $dI/dV$ calculated for a three-dimensional parabolic band of bandwidth $2$eV (crossing the Fermi energy at $|k|=0.866 \pi/a_0$), Hybridization strength $V=0.1$eV, the tunneling amplitude ratio $t_f/t_c=0.15$ and f-level energy  $\e_f=-10$eV \emph{without any finite level width} (dashed line). The shape of the curve
is similar to what has been found in Refs.~\cite{Maltseva,Morr} within
pseudoparticle mean field theory of the Kondo lattice model. These results
are nowhere near what is observed in experiments on strongly correlated
heavy fermion compounds. We stress again the main conclusion from this section: in the absence of finite f-level lifetime and breaking of translation invariance, the DOS
cannot exhibit a Fano lineshape. Therefore, understanding the origin of either finite QP lifetime or a breaking of translational invariance of the system by the influence of the tip on the surface is crucial to understanding the Fano lattice observed in recent experiments.


\sec{Tunneling into heavy fermion compounds} A microscopic Hamiltonian believed to capture the physics of heavy fermion
compounds is the lattice Anderson model

\begin{equation}
H_{AL}=H_{wc}+U\sum_{i}n_{fi\uparrow }n_{fi\downarrow }~~,\label{AndersonLattice}
\end{equation}
where $U$ is the Coulomb interaction matrix element, assumed to be much larger
than the conduction band width.

As above the single particle Green's functions as well as the self-energy
form $2\times 2$ matrices in the conduction electron (c) , f-electron (f)
space. It is worth noting that the only self energy element is $\Sigma
_{ff}(\omega )$ . Expanding $\Sigma _{ff}(\omega )$ about the Fermi energy $%
\omega =0$ (neglecting its weak momentum dependence), we end up with a
QP description in terms of the QP weight factor $z$ , the
shifted position of the f-level $\widetilde{\epsilon }_{f}$ and the QP line
width (the inverse of the QP life time) $\gamma $ , which are defined in
terms of $\Sigma _{ff}(\omega )$ by
\begin{eqnarray}
z^{-1} &=&1-\frac{\partial }{\partial \omega }\Re\Sigma _{ff}(\omega
)|_{0}, \\
\widetilde{\epsilon }_{f} &=&z[\epsilon _{f}+\Re\Sigma _{ff}(0)] \\
\gamma (E_{\mathbf{k}}) &=&z\Im\Sigma _{ff}(E_{\mathbf{k}}),
\end{eqnarray}%
The complex-valued QP energies of the two hybridized bands are
defined by

\beqa
\xi _{\mathbf{k}}^{+,-}&=&\frac{1}{2}\{\widetilde{\epsilon }%
_{f}+\epsilon _{\mathbf{k}}-i\gamma \pm \sqrt{(\widetilde{\epsilon }%
_{f}-\epsilon _{\mathbf{k}}-i\gamma )^{2}+4zV^{2}}\nonum &=& E_{\mathbf{k}}-i\Gamma _{%
\mathbf{k}}
\eeqa
The hybridization gap follows (neglecting the QP width) as $\Delta =zV^{2}(%
\frac{1}{|\epsilon _{\mathbf{k=0}}|}+\frac{1}{|\epsilon _{\mathbf{k=G}}|})$
, where $\mathbf{G}$ is a vector at the edge of the Brillouin zone. In the
heavy fermion regime, assuming that the Fermi energy ($E_{\mathbf{k}_{F}}=0$%
) intersects the lower band and observing $z^{-1}>>1$ , and $\epsilon _{%
\mathbf{k}_{F}}>>|\widetilde{\epsilon }_{f}|,\sqrt{z}|V_{\mathbf{k}}|$ one
may expand $E_{\mathbf{k}}$ about the Fermi momentum $\mathbf{k}_{F}$ to get
$E_{\mathbf{k}}=(k-k_{F})v_{F}\frac{m}{m^{\ast }(E_{\mathbf{k}})}$ , where
we defined the effective mass ratio $\frac{m}{m^{\ast }(E_{\mathbf{k}})}=z(%
\frac{V}{\epsilon _{k_{F}}})^{2}$ . Here $\epsilon _{k_{F}}$ is the value of
the bare conduction band energy at the Fermi surface, which is assumed to be
close to the upper band edge , so that $\epsilon _{k_{F}}\approx W/2$, where
$W$ is the conduction band width. Approaching the heavy fermion regime from
high temperatures the renormalization is observed to occur below a scale $%
T_{K}$ , usually called the lattice Kondo temperature. We assume$\ T_{K}$ to
be approximately equal to the Kondo temperature of a single Kondo ion.

It follows from the Pauli principle that the QP width tends to
zero in the limit temperature $T$ and excitation energy $E_{\mathbf{k}}$ $%
\rightarrow 0$ , as given by
\begin{equation}
\Gamma (E_{\mathbf{k}})=A[E_{\mathbf{k}}^{2}+(\pi T)^{2}]<<E_{\mathbf{k}}
\end{equation}%
As long as this condition is satisfied, one may employ the formalism of
Fermi liquid theory. There are indications from both , theory and experiment (for a recent determination of $\Gamma$  from ESR data see \cite{AW})
that the prefactor scales with the inverse Kondo temperature as $A\approx
1/T_{K}$. At the energy scale $T_{K}$ we therefore have $\Gamma
(T_{K})\approx T_{K}$ . Beyond that scale the lattice coherence is
suppressed by inelastic processes and the metal behaves like a crystal of
independent Kondo ions. The coherence scale of the single Kondo ion is set
by the local spin relaxation time, which is known to be $\Gamma _{s}\approx
T_{K}$ at energy/temperature less or equal to $T_{K}$ . At higher energy the
relaxation rate is given by $\Gamma _{s}(E)\approx g^{2}(E)|E|$, where $%
g(E)\approx 1/\ln (|E|/T_{K})$, $E>2T_{K}$ . At temperatures $T<<T_{K}$ one
may therefore approximate the QP width by

\begin{eqnarray}
\Gamma (E_{\mathbf{k}}) &=&E_{\mathbf{k}}^{2}/T_{K}\text{ , \ \ \ \ }E_{%
\mathbf{k}}<T_{K} \nonum
\Gamma (E_{\mathbf{k}}) &=&|E_{\mathbf{k}}|/[1+\ln (|E|/T_{K})]^{2}\text{ \
, \ }E_{\mathbf{k}}>T_{K}~~.\label{QPwidth}
\end{eqnarray}

The solid points in Fig.~\ref{figFano} denote the DOS calculated in the presence of the finite  QP width of Eq.~\ref{QPwidth}, taking $T_K=130$K \cite{Aynajian}. The solid black line is a fit to the Fano lineshape , given by \cite{Fano} $\rho_{\textrm{Fano}}=\frac{\left( q+\tilde{\e}\right)^2}{\tilde{\e}^2+1}$ where $\tilde{\e}=\frac{\e-\e_0}{\Gamma}$. The fit yields  $q=1.25$ and $\Gamma=20.7$ meV, in good agreement with the values obtained in experiment \cite{Schmidt,Aynajian}. For a qualitative comparison, in the inset we plot the experimental Fano lineshape of Ref.~\cite{Schmidt}. From  calculations of the DOS and fitting to a Fano lineshape with various parameters, we find the relation $\Gamma \sim 2T_K$ to hold for a wide range of parameter values.

In the experiment \cite{Schmidt} the Fano width $\Gamma$ also varies with the STM tip position, although the modulation is smaller than that of the Fano asymmetry parameter $q$. While for the latter the origin of the modulation is clear (since the tunneling amplitudes $t_c$ and $t_f$ depend on the tip position), a modulation of $\Gamma$ is somewhat surprising. A possible interpretation is provided in the following.

\sec{Tip-induced effects} We point out two ways in which the tunneling tip may influence the properties of the surface and consequently the tunneling lineshape. First, the f-electron self-energy acquires a contribution originating from tunneling into the tip,  $\Im{\Sigma_{ff,\textrm{tip}}(\omega)}\propto |t_f|^2 N_t(\omega)$, resulting in a modulation of the QP width with tip position. The fact that this is only a part of the self-energy accounts for the relatively weak modulation in $\Gamma$.

\begin{figure}
\includegraphics[width=8.5truecm]{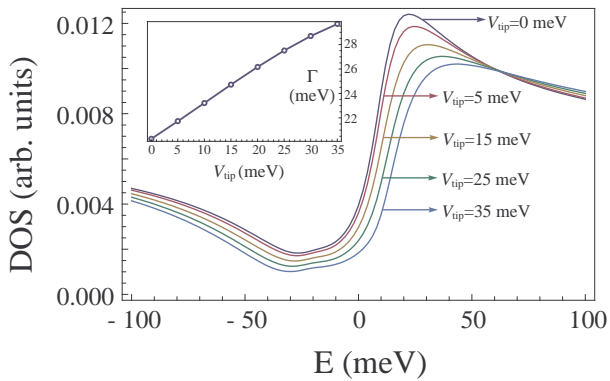}
\caption{Density of states of the two-band model Eq.~\ref{HybridHamiltonian} including the QP width of Eq.~\ref{QPwidth} and the the effect of the tip-induced hybridization, for various values of the tip potential $V_{tip}$ . The inset shows the variation of the Fano width with increasing $V_{tip}$ }.
\label{Fano_Vtip}
\end{figure}

 A second and potentially more important effect is related to the local potential generated by the tip at the surface underneath \cite{Cheng}. This potential acts like an additional local hybridization between f- and c-states, and will give rise to Kondo impurity type behavior. More specifically, in addition to the interaction-induced broadening, there will be a tip-induced \emph{local} scattering term added to the Hamiltonian of Eq.~\ref{HybridHamiltonian}, of the form
\beqa H_{tip}&=&V_{tip} c^{\dagger}_{{\bf \textrm{r}}=0} f_{{\bf \textrm{r}}=0}+h.c. \nonum
&=& V_{tip} \sum_{k,k'} c^{\dagger}{k}f_{k'}+h.c. \label{Htip} \eeqa where $\textrm{r}=0$ is the tip position and $V_{tip}$ is the hybridization strength, which will depend on the tip position as well as on other parameters such as tip height above the surface and operating voltage. It is simple to show \cite{Dubi} that this kind of hybridization, which breaks translation invariance, directly leads to a Fano lineshape even in the absence of c-f hybridization. In the presence of the hybridization and the intrinsic broadening described above, the DOS can be evaluated by solving the Dyson equation in the c-f :
\beq \hat{G}_{kk'}=\hat{G}^0_{k} \delta (k,k')+\hat{G}^0_k\cdot \hat{V}_{tip}\cdot \left(1-\hat{V}_{tip} \cdot \hat{\chi} \right)^{-1} \cdot\hat{G}^0_{k'}~~, \eeq where $\hat{G}^0_k$ is the $2\times 2$ Green's function matrix $\hat{\chi}=\sum_k \hat{G}^0_{k}$ and $\hat{V}_{tip}=V_{tip} \left(
                                                                           \begin{array}{cc}
                                                                             0 & 1 \\
                                                                             1 & 0 \\
                                                                           \end{array}
                                                                         \right)
$. In Fig.~\ref{Fano_Vtip} we plot the DOS calculated with $V_{tip}=0,5,15,25,35$ meV (shown by arrows in the figure). The lineshape seems to be hardly affected, however, a fit to the Fano lineshape shows that while the Fano asymmetry parameter $q$ hardly changes (within $10 \%$), the Fano width $\Gamma$, depicted in the inset as a function of $V_{tip}$, changes by $150\%$, as seen in experiment. We therefore conclude that while the dominant cause for modulation of $q$ is the modulation of the tunneling amplitudes, the modulation in $\Gamma$ may be due to modulation of the tip-induced local scattering.

\sec{Conclusion}
In this paper we offer an interpretation of the recent observation of a lattice periodic pattern of Fano-shaped tunneling spectra of heavy fermion metals, which appears to be very similar to what is observed at Kondo impurity atoms on the surface. First we demonstrate for the case of weakly correlated hybridized  bands that  the tunneling spectra reflect the local density of states, which  does not show a Fano lineshape in  general. We propose the intrinsic  \emph{correlation induced} quasi-particle line broadening as the origin of the Fano lineshape observed in tunneling experiments on strongly correlated metals. We model the quasi-particle energy width phenomenologically by using Fermi liquid theory and Kondo physics. In fact the broadening is related to the Kondo temperature via Eq.~\ref{QPwidth}, and yields a Fano linewidth $\Gamma\sim 2T_k$, in agreement with recent experiments.

In addition, we address the modulation in the Fano width $\Gamma$ observed experimentally. While the lattice periodic modulation of the Fano asymmetry parameter $q$ is immediately understood as a consequence of the modulation of the tunneling amplitudes between tip and f- or c-orbitals, the modulation of the width $\Gamma$ requires a different interpretation, possibly related to the influence of the tip on the properties of the metal beneath it. Specifically, we show that a tip-induced local hybridization may generate such a modulation. A direct consequence would be that the Fano parameters $q$ and $\Gamma$ would depend on STM tip parameters (height, voltage etc.). A dependence of $q$ on tip parameters is rather expected, since is depends on the tunneling amplitudes $t_c$ and $t_f$, which are unlikely to change in the same way with changing tip parameters, hence $t_f/t_c$ should depend on the tip parameters. However, a dependence of $\Gamma$ on tip parameters would be surprising if $\Gamma$ is a purely intrinsic quantity, and would indicate that indeed the STM tip effect in non negligible. These predictions can be directly tested within current experimental setups.

\begin{acknowledgements}
We acknowledge stimulating discussions with C. Seamus Davis, M. Graf, J. Fransson and P. Coleman.
PW acknowledges partial support by the Theory Division of Los Alamos Laboratory, the Aspen Center for Physics and the DFG research unit "Quantum phase transitions". AVB and YD acknowledge support by US BES and UCOP-TR-01.

\end{acknowledgements}



\begin{thebibliography}{99}

\bibitem{Schmidt}
A. Schmidt \etal,, Nature {\bf 465}, 570 (2010).

\bibitem{Aynajian}
P. Aynajian \etal, PNAS {\bf 107}, 10383 (2010).

\bibitem{Kondo}
J. Kondo, Prog. Theor. Phys. {\bf 32}, 37 (1961).

\bibitem{Hewson}
A.C. Hewson, \textit{The Kondo Problem to Heavy Fermions} (Cambridge University Press, 1993).

\bibitem{exp}
M. B. Maple et al., J. Low Temp. Phys.{\bf 99}, 223 (1995); A. Schr\"{o}der et al., Nature (London) {\bf 407}, 351 (2000); J. Custers et al., Nature (London) {\bf 424}, 524 (2003); P. Gegenwart, Q. Si, and F. Steglich, Nature Phys. {\bf 4}, 186 (2008); for reviews see: G. R. Stewart, Rev. Mod. Phys. {\bf 73}, 797 (2001);  H. v. Löhneysen, Rev. Mod. Phys. {\bf 79}, 1015 (2007).

\bibitem{Doniach}
S. Doniach, Physica (Amsterdam) {\bf 91}, 231 (1977).

\bibitem{theory}
Q. M. Si et al., Nature (London) {\bf 413}, 804 (2001); P. Coleman et al., J. Phys.Condens. Matter {\bf 13}, R723 (2001);  T. Senthil, S. Sachdev, and M. Vojta, Phys. Rev. Lett. {\bf 90}, 216403 (2003); I. Paul, C. Pepin, and M. R. Norman,  Phys. Rev. Lett. {\bf 98}, 026402 (2007); Y.-F..Yang et al., Nature (London) {\bf 454}, 524 (2008).

\bibitem{Maltseva}
M. Maltseva, M. Dzero, and P. Coleman, Phys. Rev. Lett. {\bf 103}, 206402 (2009).

\bibitem{Morr}
J. Figgins and D. K. Morr, Phys. Rev. Lett. {\bf 104}, 187202 (2010).

\bibitem{Hershfield}
S. Hershfield, J. H. Davies, and J. W. Wilkins, Phys. Rev. Lett. {\bf 67}, 3720 (1991).

\bibitem{Ujsaghy}
O. \'{U}js\'{a}ghy, J. Kroha, L. Szunyogh, and A. Zawadowski, Phys. Rev. Lett. {\bf 85}, 2557 (2000).

\bibitem{Plihal}
M. Plihal and J. W. Gadzuk, Phys. Rev. B {\bf 63}, 085404 (2001).

\bibitem{Merino}
J. Merino and O. Gunnarsson, Phys. Rev. B 69, 115404 (2004).

\bibitem{Madhavan}
V. Madhavan, W. Chen, T. Jamneala, M. F. Crommie, and Ned S. Wingreen, Phys. Rev. B {\bf 64}, 165412 (2001).

\bibitem{Nagaoka}
K. Nagaoka, T. Jamneala, M. Grobis, and M. F. Crommie, Phys. Rev. Lett. {\bf 88}, 077205 (2002).

\bibitem{Yang}
Y. -F. Yang,  \prb {\bf 79}, 241107 (2009).

\bibitem{Fogelstrom}
M. Fogelstr\"{o}m \etal, cond-mat/1004.1882 (unpublished).


\bibitem{Fano}
U. Fano, Phys. Rev. {\bf 124}, 1866 (1961).

\bibitem{Cheng}
P. Cheng \etal, cond-mat/1001.3220 (unpublished).

\bibitem{Dubi}
Y. Dubi and A. V. Balatsky, submitted.

\bibitem{AW}
E. Abrahams, and P. W\"{o}lfle, Phys. Rev. B {\bf 78}, 104423 (2008); P. W\"{o}lfle, and E. Abrahams, Phys. Rev. B {\bf 80}, 235112 (2009)
%
%
%
\end{thebibliography}
\end{document}